\documentstyle[12pt]{article}
\begin{document}
\newcommand{\beq} {\begin{equation}}
\newcommand{\eeq} {\end{equation}}
\newcommand{\ep} {\epsilon^{\mu\nu}}

\title
{Fermionic Coset Models as Topological Models}
\author{
G.L.\ Rossini\thanks{Partially supported by CONICET, Argentina}
\rule{0cm}{1.5cm}  and
F.A.\ Schaposnik\thanks{Fellow of the J.S.Guggenheim Memorial
Foundation, USA. Investigador CICBA, Argentina}\\
Departamento de F\'\i sica, Universidad Nacional de La
Plata\thanks{Postal
address: C.C. 67, (1900) La Plata, Argentina} \\
Argentina}
\date{}
\maketitle

\def\thepage{\protect\raisebox{0ex}{\ } La Plata 92-08}
\thispagestyle{headings}
\markright{\thepage}

\begin{abstract}
By considering the fermionic realization of $G/H$ coset models, we show
that the partition function for the $U(1)/U(1)$ model
defines a Topological Quantum Field
Theory and coincides with that for a 2-dimensional Abelian BF system.
In the non-Abelian case, we prove the topological character of
$G/G$ coset models by explicit computation, also
finding a natural extension of 2-dimensional BF systems with non-Abelian
symmetry.
\end{abstract}
\newpage
\pagenumbering{arabic}

In the last few years coset models \cite{Pr}-\cite{GKO} raised much
interest in the study of conformally invariant two-dimensional theories
particularly in connection with String theories
and with Statistical Mechanics models.\cite{Rev}

$G/H$ coset models can be realized by gauging a subgroup $H$ of: (i) a
Wess-Zumino-Witten (WZW) model with the basic field taking values
on a Lie group G \cite{K}-\cite{Schn} or, alternatively, (ii) a
free fermionic model
with fermions in the fundamental representation of $G$
\cite{BH}-\cite{cecilia}.

Recently, Witten \cite{Witten} analysed the holomorphic factorization
of $G/H$ models (in its bosonic realization) showing in particular
that the $G/G$ model defines a Topological field
theory (i.e. a quantum field theory with metric independent partition
function). We discuss this issue in the present note, by considering
the {\it fermionic} realization of coset models.

The coset construction based on fermionic models goes as follows
\cite{cecilia}. One starts with two-dimensional free
Dirac fermions in the fundamental representation
of $G$, with Lagrangian:

\beq
L_0 = \bar \psi i\!\!\not\!\partial \psi \label{1}
\eeq
Calling $t^a$ the generators for $H \subset G$ one constructs the
associated currents:

\beq
j_{\mu}^a = \bar\psi t^a\gamma_{\mu}\psi \label{2}
\eeq
Then, one imposes the condition that physical states $\vert phys>$
are singlet under these currents:

\beq
j_{\mu}^a \vert phys> = 0 \label{3}
\eeq
This is achieved in the path-integral formulation by introducing Lagrange
multipliers $A_{\mu}^a$ which play the role of gauge fields in the
Lie algebra of $H$. The partition function for the resulting constrained
model reads:

\beq
Z_{G/H} = \int D\bar\psi D\psi \prod_a \delta (J_{\mu}^a)
exp[-\int_M \sqrt g d^2x L_0]  \label{4}
\eeq
or
\beq
Z_{G/H} = \int DA_{\mu} D\bar\psi D\psi exp[-\int_M \sqrt g d^2x
\bar\psi (i\!\!\not\!\partial + \not\!\!A)\psi] \label{5}
\eeq
with $\sqrt g = (det\,g_{\mu\nu})^{\frac{1}{2}}$ and $g_{\mu\nu}$ a
metric on the two-dimensional manifold $M$.
One easily verifies that the constrained model defined
by (\ref{5}) corresponds to a coset model with Virasoro central charge
\cite{cecilia}:

\beq
c_{G/H} = c_{G} - c_{H} \label{6}
\eeq

Of particular interest are $G/G$ models which have
been shown to be (in their bosonic formulation) topological field
theories
\cite {Witten}. Note that the fermionic version of the $G/G$ model given
by eq.(\ref{5}) is just:

\beq
Z_{G/G} = \int DA_{\mu} det (i\!\!\not\!\partial + \not\!\!A) \label{7}
\eeq
with $A_{\mu}$ taking values in the Lie algebra of $G$ and hence
$Z_{G/G}$
corresponds to the $QCD_2$ partition function in the infinite coupling
constant limit. Of course, an
appropriate gauge-fixing is necessary in (\ref{7}).

As stated above, it is the purpose of this note to study
$G/G$ coset models. For the sake of clarity, we shall first consider
the $G=U(1)$ case and then consider the non-Abelian extension.
In the $U(1)$ case, the fermionic determinant in (\ref{7}) takes the
form \cite{Schw}:

\beq
det (i\!\!\not\!\partial + \not\!\!A) = exp[-\frac{1}{4\pi}\int_M d^2x
\ep F_{\mu\nu} \varphi] \times det\; i\!\!\not\!\partial ,\label{8}
\eeq
where $A^{\mu}$ and $\varphi$ are related through the decomposition:

\beq
A^{\mu} = \frac{1}{\sqrt g} \ep \partial_{\nu} \varphi +
\partial^{\mu}\eta
\label{9}
\eeq
and $\epsilon ^{01} = -\epsilon^{10} = 1$. Note that:

\beq
\frac{1}{2\sqrt g} \ep F_{\mu\nu} = -\Box \varphi \label{10}
\eeq

Now, in order to linearize the dependence
of the fermionic determinant (\ref{8})
on $A_{\mu}$, we introduce a scalar
field $\phi$ through the identity:
\beq
det (i\!\!\not\!\partial + \not\!\!A) = \frac{det\; i\!\!\not\!\partial}
{det^{-\frac{1}{2}}\Box} \int D\phi
\exp(-\frac{1}{8\pi}\int_M \phi\Box\phi\sqrt{g}
d^2x -\frac{1}{4\pi}\int_M  \ep F_{\mu\nu} \phi d^2x)
\label{11}
\eeq
integration and
Determinants on the r.h.s. of eq.(\ref{11}) coincide
with the partition function for free Dirac fermions and with the one for
free bosons, both in the presence of a background metric. Due to the
boson-fermion
connection in two dimensions, they define equivalent theories.
Without lack of generality one can choose $g_{\mu\nu}$ as a conformally
flat metric, $g_{\mu\nu}=\exp(\sigma)\delta_{\mu\nu}$.
One can then show that \cite{dH}:

\beq
{det^{-\frac{1}{2}}\Box} = {det\; i\!\!\not\!\partial} =
\exp(S_L[\sigma])
\label{12}
\eeq
where $S_L$ is the Liouville action for the scalar field $\sigma$.
(We have disregarded in eq.(\ref{12}) metric independent constants).
Hence, their contribution in eq.(\ref{11}) cancels out.

With this, the partition function for the $U(1)/U(1)$ coset model
can be written as
\beq
Z_{U(1)/U(1)}=\int DA_{\mu} D\phi D\bar{c} Dc D\pi exp(-S_Q) \label {13}
\eeq
with
\beq
S_Q= \frac{1}{8\pi}\int_M \phi\Box\phi\sqrt{g} d^2x
+\frac{1}{4\pi}\int_M  \ep F_{\mu\nu} \phi d^2x +
\int_M \{ Q,\bar{c}G[A_{\mu}]\} \sqrt{g} d^2x
\label{14}
\eeq
where the gauge fixing term corresponds to some gauge condition $G[A]=0$
and  BRST transformations $\{Q,~\}$ are defined as:

\beq
\begin{array}{ll}
\{Q,A_{\mu}\} = -\partial_{\mu}c & \{Q,c\} = 0 \\
\{Q,\bar c\} = \pi & \{Q,\pi\} = 0\\
\{Q,\phi\}=0
\end{array}
\label {15}
\eeq
Here $c$ and $\bar{c}$ are ghost fields and $\pi$ is a Lagrange
multiplier.

For simplicity of the arguments below, we choose the axial gauge
which in light cone coordinates reads:
\beq
G[A_{\mu}]\equiv A_- =0
\label{16}
\eeq
so that the gauge fixing term becomes
\beq
\{Q,\bar{c}_+A_-\}=\pi_+ A_- - \bar{c}_+\partial_-c
\label{17}
\eeq

It is now easy to prove by explicit computation
that $Z_{U(1)/U(1)}$ does not depend on the metric.
Indeed, the ghost field integration gives as Fadeev-Popov
determinant $det\,\partial_-$. As the $\pi$ integration implements the
gauge condition yielding $\delta(A_-)$, the $A_-$ integration is trivial
so that one gets
\beq
Z_{U(1)/U(1)}=det\,\partial_-\int DA_+ D\phi \exp(-\frac{1}{8\pi}\int_M
(\phi\Box\phi\sqrt{g}-4\partial_-A_+\phi)d^2x)
\label{18}
\eeq
Now the $A_+$ integration imposes the constraint $\partial_-\phi=0$
and hence the Laplacian in the exponential vanishes. We have finally
\beq
Z_{U(1)/U(1)}=det\,\partial_-\int D\phi \delta(\partial_-\phi)=1
\label{19}
\eeq
We have then proved the metric independence of $Z_{U(1)/U(1)}$, i.e. the
topological character of the $U(1)/U(1)$ coset model.
Of course, given a theory defined on a manifold $M$  with a
{\it fixed metric},
the corresponding partition function is a number which can be normalized
to 1. What eq.(\ref{19}) means is that this normalization does not change
when the metric is varied.
(The extension of our proof to an arbitrary gauge condition is trivial.)

The same result can be more elegantly obtained by connecting
the partition function in eqs.(\ref{13}-\ref{14}) with that of an Abelian
BF system\footnote{We thank M.\ Henneaux for suggesting this connection.}
(see \cite{Bir} and references therein). To see this, let us perform
in eq.(\ref{14}) the following change of variables
\beq
A_{\mu}\to A_{\mu}+\frac{\sqrt g}{4} \epsilon_{\mu\nu}\partial^{\nu}\phi
\label{20}
\eeq
which leaves invariant the path-integral measure in (\ref{13}). After
this change $Z_{U(1)/U(1)}$ takes the form:
\beq
S_Q=\frac{1}{4\pi}\int_M  \ep F_{\mu\nu} \phi d^2x +
\int_M \{ Q,\bar{c}\tilde{G}[A_{\mu}]\} \sqrt{g} d^2x
\label {21}
\eeq
Eq.(\ref{21}) is the quantum action corresponding to a BF system
for a scalar field $\phi$ and a gauge field $A_{\mu}$, with
classical action
\beq
S_{BF}=\frac{1}{4\pi}\int_M  \ep F_{\mu\nu} \phi d^2x
\label{22}
\eeq
and the appropriate gauge fixing (note that in the case we were working
in the Landau gauge, the gauge fixing functional $G[A_{\mu}]=
\frac{1}{\sqrt{g}}\partial_{\mu}(\sqrt{g}A^{\mu})$ would remain unchanged
after performing the change of variables).
Then, taking into account the invariance
of the path-integral measure $DA_{\mu}$, eq.(\ref{13})
becomes the partition function for the BF system:
\beq
Z_{U(1)/U(1)}=Z_{BF}
\label{23}
\eeq

It is well known \cite{Bir}
that the partition function of an Abelian
BF system defined on an $n$-dimensional manifold $M_n$ is a topological
invariant; moreover, it gives some power of the Ray-Singer torsion of
$M_n$, which is one in even-dimensional manifolds.
In fact, one can prove this last result just by following analogous
steps of those leading from eq.(\ref{13}) to eq.(\ref{19}).
Thus, the relation (\ref{23}) is
consistent with our result in eq.(\ref{19}).

\vspace{5mm}
Let us now extend our derivation to the case of a  non-Abelian
group $G$ (which will be taken as a compact Lie group).
The fermion determinant appearing
in the partition function $Z_{G/G}$ (eq.(\ref{7})) can be written,
in the $A_- = 0$ gauge, in the form \cite{det}:

\beq
det (i\!\!\not\!\partial + \not\!\!A) = exp( I[h]) \times
det\; i\!\!\not\!\partial
\label{24}
\eeq
with $h$ a G-valued field related to $A_+$ through:
\beq
A_+ = h^{-1}\partial_+ h
\label{25}
\eeq
and $I[h]$ the WZW action:

\beq
I[h] = \frac{1}{8\pi}tr\int_M d^2x \sqrt{g} \partial^{\mu}h^{-1}
\partial_{\mu}h
+ \frac{1}{12\pi}tr\int_B d^3y \epsilon^{ijk} h^{-1}\partial_i h
h^{-1}\partial_j h h^{-1}\partial_k h \label{26}
\eeq
Here $B$ is a three dimensional manifold such that $\partial B = M$.
The third coordinate in $B$, which we call $t$, will be taken as usual
as $t\in [0,1]$. Then $h(x,t)$ is an extension of $h(x)$ over $B$ such
that $h(x,1)=h(x)$ and \linebreak $h(x,0)=1$, the unit element of G.

The dependence of the fermionic determinant (\ref{24}) on $A_{\mu}$ can
be linearized, as in the Abelian case, by introducing a G-valued scalar
field $g$. Indeed, using the Polyakov-Wiegmann identity \cite{PW}
\beq
I[hg]=I[g]+I[h]-\frac{1}{4\pi}tr\int_M h^{-1}\partial_+h
\partial_-gg^{-1}d^2x
\label{27}
\eeq
one can easily see that
\beq
det (i\!\!\not\!\partial + \not\!\!A) =\frac{det\; i\!\!\not\!\partial}
{Z_{WZW}}\int Dg\,\exp(-I[g]+\frac{1}{4\pi}
tr\int_M (A_+\partial_-gg^{-1})
d^2x)
\label{28}
\eeq
where
\beq
Z_{WZW}=\int Du\,\exp(-I[u])
\label{29}
\eeq
is the WZW partition function
and the argument in the exponential in eq.(\ref{28}) is minus the gauged
WZW action in the $A_-=0$ gauge.

The determinant appearing in the r.h.s. of eq.(\ref{28})
corresponds to the partition function for  free Dirac fermions in the
fundamental representation of G
while $Z_{WZW}$ is the partition function for the equivalent bosonic
theory. One can again show \cite{BGRN}, exploiting the (non-Abelian)
boson-fermion equivalence in two dimensions, that these partition
functions are identical:
\beq
det\; i\!\!\not\!\partial = Z_{WZW}
\label{33}
\eeq
Putting all this together we have for the
$G/G$ coset model partition function:
\beq
Z_{G/G}=\int DA_+DA_-DgD\bar{c}_+DcD\pi_+\,\exp(-S_Q)
\label{34}
\eeq
with
\beq
S_Q=I[g] - \frac{1}{4\pi}tr\int_M (A_+\partial_-gg^{-1}) d^2x +
tr\int_M\{Q,\bar{c}_+A_-\}\sqrt{g}d^2x
\label{35}
\eeq
Here the gauge fixing corresponds to the gauge condition $A_-=0$ and BRST
transformations are defined as

\beq
\begin{array}{ll}
\{Q,A_{\mu}\} = -D_{\mu}c & \{Q,c\} = \frac{1}{2}[c,c] \\
\{Q,\bar{c}_+\} = \pi_+ & \{Q,\pi_+\} = 0\\
\{Q,g\}=-[g,c]
\end{array}
\label {36}
\eeq
with $\bar{c}_+$ and $c$ ghost fields and $\pi_+$ a Lagrange multiplier,
all of them taking values in the Lie algebra of G. The explicit form of
the gauge fixing term in eq.(\ref{35}) is
\beq
\int_M\{Q,\bar{c}_+A_-\}\sqrt{g}d^2x =
\int_M(\pi_+A_- - \bar{c}_+D_-c)\sqrt{g}d^2x
\label{37}
\eeq
As in the Abelian case, we can now perform the explicit computation of
eq.(\ref{34}). The ghost field integration yields the Fadeev-Popov
determinant $det\,D_-$, while the $\pi_+$ integration implements the
gauge condition $\delta(A_-)$. The $A_-$
integration then sets $det\,D_-=det\,\partial_-$ and one then ends with
\beq
Z_{G/G}=\int DA_+Dg\,det\,\partial_-\exp(-I[g]+\frac{1}{4\pi}tr\int_M
A_+\partial_-gg^{-1}d^2x)
\label{38}
\eeq
We see that the $A_+$ integration in eq.(\ref{38}) imposes
the constraint $\partial_-gg^{-1}=0$ for
each point on the manifold $M$. Moreover, one can find
an appropriate extension of $g(x)$ over
$B$ such that $\partial_-gg^{-1}=0$ for every point in $B$.
With this eq.(\ref{38}) becomes
\beq
Z_{G/G}=det\,\partial_-\int Dg\,\delta(\partial_-gg^{-1})
\label{39}
\eeq
The integration over $g$ is most easily performed by writing
$g=\exp(\alpha)$
($\alpha$ in the Lie algebra of G) and integrating over $\alpha$. Using
\beq
\frac{\delta (\partial_-gg^{-1})}{\delta\alpha}
\vert_{\partial_-gg^{-1}=0}=\partial_-
\label{40}
\eeq
we then finally get
\beq
Z_{G/G}=1
\label{41}
\eeq
Hence, as in the Abelian case, we have
proved that $Z_{G/G}$ is metric independent thus defining
a topological quantum field theory
(This proof should be extended to an arbitrary gauge without
difficulty).

It is important to stress at this point that for $G/H$ coset models with
$H \neq G$, an identity analogous to (\ref{33}) is not valid. Indeed,
for $H \neq G$, $g$ should belong to subgroup $H$ and $A_{\mu}$ to its
Lie algebra,
while fermions should still be in the fundamental representation of $G$.
Then,
following the steps described above, one should arrive to a relation of
the form (\ref{33}) with $det\: i\!\!\!\,\not\!\!\partial$
still being the partition function for free Dirac fermions in the
fundamental representation of $G$ while $Z_{WZW}$ would correspond to
a partition function of $H$-valued WZW
fields. Hence, these two partition functions would not cancel each other
as they do for $H=G$ and $Z_{G/H}$ would be {\it metric-dependent}.

Let us now discuss the non-Abelian analogue of the steps leading to
the equivalence between the Abelian coset model and a BF system.
After some algebra, $S_Q$ in eq.(\ref{35}) can be written as
\begin{eqnarray}
\lefteqn{S_Q=\frac{1}{4\pi}\int_M d^2x\int_0^1 dt\,
tr[g^{-1}(x,t)\partial_tg(x,t)} \nonumber \\
& &\partial_-(g^{-1}(x,t)A_+(x)g(x,t)+
g^{-1}(x,t)\partial_+g(x,t))]+ \nonumber \\
& &+tr\int_M \{Q,\bar{c}_+A_-\}\sqrt{g}d^2x
\label{42}
\end{eqnarray}
Now, defining a field $\tilde{A}_+(x,t)$ over $B$ as
\beq
\tilde{A}_+(x,t)=g^{-1}(x,t)A_+(x)g(x,t) + g^{-1}(x,t)\partial_+g(x,t)
\label{43}
\eeq
(compare with the transformation in eq.(\ref{20}) for the Abelian case,
setting $g(x,t)=\exp(it\phi(x))$ and noting that
$\partial_+=\frac{1}{\sqrt{g}}\partial^-$ in a conformally flat metric)
we can write
\begin{eqnarray}
S_Q=\frac{1}{4\pi}\int_M d^2x\int_0^1 dt\,
tr[\partial_-\tilde{A}_+(x,t)
g^{-1}(x,t)\partial_tg(x,t)]+ \nonumber \\
+tr\int_M \{Q,\bar{c}_+A_-\}\sqrt{g}d^2x
\label{44}
\end{eqnarray}
Note that, as in the WZW model, though $\tilde{A}_+(x,t)$ appears in
the first
integral in eq.(\ref{44}), $S_Q$ is a functional of $\tilde{A}_+$
on $M$, i.e. a functional of $\tilde{A}_+(x,1)$. So we can
change variables from $A_+(x)$ to $\tilde{A}_+(x,1)$ in the
path-integral (\ref{34}). From eq.(\ref{43}) we see that the
Jacobian associated with this change is trivial, and hence we get
\beq
Z_{G/G}=\int D\tilde{A}_+ DA_-DgD\bar{c}_+ Dc D\pi_+ \exp(-S_Q)
\label{45}
\eeq
with $S_Q$ given by eq.(\ref{44}). Note that in  terms of the
integration variable $\tilde{A}_+(x,1)$, we can write
\beq
\tilde{A}_+(x,t)=u^{-1}(x,t)\tilde{A}_+(x,1)u(x,t)
+u^{-1}(x,t)\partial_+u(x,t)
\label{46}
\eeq
with $u(x,t)=g^{-1}(x,1) g(x,t)$. Comparing expression (\ref{44})
with the one
obtained for the Abelian case (eq.(\ref{21})), we see that it is
sensible to write
\beq
S_Q=\tilde{S}_{BF}+ tr\int_M \{Q,\bar{c}_+A_-\}\sqrt{g}d^2x
\label{47}
\eeq
with
\beq
\tilde{S}_{BF}=\frac{1}{4\pi}\int_M d^2x\int_0^1 dt\,
tr[\partial_-\tilde{A}_+(x,t) g^{-1}(x,t)\partial_tg(x,t)]
\label{48}
\eeq
representing the natural extension of the 2-dimensional Abelian BF system
defined by action (\ref{22})
to the non-Abelian case. With this interpretation not only we have again
\beq
Z_{G/G}=\tilde{Z}_{BF}=1
\label{49}
\eeq
but also parallel the route followed when one extends the bosonization
recipe from the Abelian to the non-Abelian case. Both in the
bosonization procedure
and in our proof above, the basic objects
in the non-Abelian case are constructed from group elements
$g$ and one needs an extension of the original 2-dimensional manifold
$M$ to the ball $B$ in order to have a closed expression for the
Lagrangians (eqs.
(\ref{26}) and (\ref{48})).
One can then conclude that the non-Abelian version of BF systems
discussed in the literature, consisting in writing an action like in
eq.(\ref{22}) but with $\phi$ and $F_{\mu\nu}$ in the Lie algebra of G,
is just the counterpart, when studying BF systems, of taking
$\int \partial_{\mu}\phi^a \partial^{\mu}\phi^ad^2x$ as the bosonized
form of a 2-dimensional fermionic theory with symmetry group G.
As it is well-known, a major limitation of this bosonization
procedure is, however, that the non-Abelian symmetry is not preserved
by the bosonization. In view of the connection of fermionic
and bosonic versions of coset models,
we then prefer to consider $\tilde{S}_{BF}$ as defined in eq.(\ref{48})
as the natural non-Abelian extension of BF systems.

\vspace{5mm}
Let us end this work by
discussing  a second supersymmetry (appart
from BRST symmetry),  which can be implemented
in the $U(1)/U(1)$ model
(and presumably extended to the non-Abelian case).
For that purpose, we first choose the Landau gauge
$G[A_{\mu}]=\frac{1}{\sqrt{g}}\partial_{\mu}(\sqrt{g}A^{\mu})$
to write $S_Q$ in eq.(\ref{14}) as:
\begin{eqnarray}
\lefteqn{S_Q= \frac{1}{8\pi}\int_M \phi\Box\phi\sqrt{g} d^2x
+\frac{1}{4\pi}\int_M  \ep F_{\mu\nu} \phi d^2x +} \nonumber \\
& & +\int_M (\pi \frac{1}{\sqrt{g}}\partial_{\mu}(\sqrt{g}A^{\mu}) -
\bar{c}\Box c)\sqrt{g} d^2x
\label{sq}
\end{eqnarray}
Calling $Q^*$ the
generator associated to this second supersymmetry, and
following Soda \cite{soda} in his analysis of two-dimensional Maxwell
theory, we define transformation laws in the form:

\beq
\begin{array}{ll}
\{Q^*,A_{\mu}\} = -\sqrt{g} \epsilon_{\mu\nu}\partial^{\nu}c & \{Q^*,c\}
= 0 \\
\{Q^*,\bar c\} = -\frac{1}{2\pi} \phi & \{Q^*,b\} = 0  \\
 \{Q^*,\phi\} = 0 & ~
\end{array}
\label {s29}
\eeq

Now, we note that $S_Q$, defined in eq.(\ref{sq}), satisfies
not only $\{Q,S_Q\} = 0$ but also:

\beq
\{Q^*,S_Q\} = 0 \label{s30}
\eeq
Indeed:

\beq
\{Q^*, \frac{1}{4\pi} \int_M d^2x \ep F_{\mu\nu} \phi\} = -\frac{1}{2\pi}
\int_M d^2x \phi\ep \partial_{\mu}(\sqrt{g} \epsilon_{\nu\alpha}
\partial^{\alpha}c)
\label{s31}
\eeq
or

\beq
\{Q^*, \frac{1}{4\pi} \int_M d^2x \ep F_{\mu\nu} \phi\} = \frac{1}{2\pi}
\int_M \sqrt{g} d^2x \phi \Box c
\label{s32}
\eeq
while

\beq
\{Q^*,\int_M \sqrt{g} d^2x \bar c \Box c \}= -\frac{1}{2\pi}
\int_M \sqrt{g} d^2x \phi \Box c
\label{s33}
\eeq
and the other terms in $S_Q$ are $Q^*$-invariant separatedly.

Let us note that $S_Q$ can be written in the form:

\beq
S_Q = \frac{1}{4\pi} \int_M d^2x \ep F_{\mu\nu} \phi
+ \{Q,V\} + \{Q^*,W\}
\label{s34}
\eeq
with

\beq
\begin {array}{ll}
V = \int_M d^2x \bar c \partial_{\mu}(\sqrt{g}A^{\mu}) &
W = -\frac{1}{4}\int_M \sqrt{g}d^2x \bar c \Box \phi
\end{array}
\label {s35}
\eeq
and that all metric dependence in $S_Q$ is in the two last terms in the
r.h.s. of eq.(\ref{s34}). One then has:

\beq
\frac{1}{Z_{U(1)/U(1)}}\frac{\delta Z_{U(1)/U(1)}}{\delta g_{\mu\nu}} =
<\{Q,\frac{\delta V}{\delta g_{\mu\nu}}\}> +
<\{Q^*,\frac{\delta W}{\delta g_{\mu\nu}}\}>  \label{s36}
\eeq
The l.h.s. in eq.(\ref{s36}) is zero due to the topological character
of the model. This, together with the condition:

\beq
Q\vert phys> = 0
\label{s37}
\eeq
on physical states $\vert phys>$ means that:
\beq
<\{Q^*,\frac{\delta W}{\delta g_{\mu\nu}}\}> = 0
\label{s37a}
\eeq
It is interesting to note that in his analysis of the two-dimensional
Maxwell model, which is not in principle a topological one by itself
\cite{Ki,BT,Witten2},
Soda had to impose $Q^* \vert phys> = 0$ in
order to define a topological theory.
In contrast, in the $U(1)/U(1)$ case,
we have shown that (\ref{s37a}) holds due to the topological
character of the coset model, without imposing Soda's condition.
Let us finally mention that the steps leading to (\ref{s36}) can be
repeated using an explict invariant measure as for example Fujikawa's
measure (see \cite{CLS}).

\vspace{5mm}
In summary, we have been able to show that $G/G$ models are topological
by starting from their fermionic realization.
That is, $Z_{G/G}$ is independent of the
metric of the 2-dimensional manifold $M$ on which the model is defined.
We have also established a connection with BF systems provided in the
non-Abelian case one considers a new class of such models.

Since Topological Quantum Field Theories are characterized by observables
which depend only on the global features of $M$, it should be of interest
to study in detail correlation functions for the fermionic realization
of $G/G$ model as a way of obtaining novel
representations of global invariants. From the point of view of Quantum
Field Theory, the connection of $G/G$ models with $QCD_2$ at strong
coupling
opens a new route to the analysis of 2-dimensional Yang-Mills theory with
matter, using Topological Quantum Field Theory tools. We hope to report
on these issues elsewhere.

\vspace{1cm}
\underline{Acknowledgements} We would like to thank M.\ Henneaux for his
collaboration at the first stages of this work and most valuable
comments. F.A.S.\ whishes to acknowledge G.\ Thompson
for useful correspondence about topological aspects of coset models.
We also thank D.Cabra and E.Moreno for helpful comments. This work was
supported in part by CONICET and Fundaci\'on Antorchas (Argentina).


\begin{thebibliography}{99}
\bibitem{Pr} K.Bardakci and M.B.Halpern, Phys.Rev. {\bf D3}(1971)2943;
M.B. Hal\-pern, Phys.Rev.{\bf D4}(1971)2398.
\bibitem{GKO} P.Goddard, A.Kent and D.Olive, Phys.Lett. {\bf B152}(1985)
88; Comm. Math.Phys. {\bf 103}(1986) 105.
\bibitem{Rev} C.Itzykson, H.Saleur and J-B.Zuber, eds., Conformal
Invariance
and Applications to Statistical Mechanics, World Scientific, Singapore,
1988.

\noindent P.Ginsparg in Strings and Critical Phenomena,
E.Brezin and J.Zinn-Justin eds., Elsevier, 1989.
\bibitem{K} K.Gawedzki and A.Kupiainen, Phys.Lett. {\bf 215B}(1988)119.
\bibitem{Schn} D.Karabali, Q.-H.Park, H.J.Schnitzer and Z.Yang,
Phys.Lett. {\bf 216B} (1989)307.
\bibitem{BH} K.Bardakci, E.Rabinovici and B.Saring, Nucl.Phys. {\bf
B299}(1988)151.
\bibitem{cecilia} D.Cabra, E.Moreno and C.von Reichenbach,
Int.Jour.Mod.Phys. {\bf A5} (1990)2313.
\bibitem{Witten} E.Witten, Commun.Math.Phys.{\bf 144} (1992)189.
\bibitem{Schw} J.Schwinger, Phys.Rev.{\bf 128} (1962)2425.
R.Roskies and F.A. Schaposnik, Phys.Rev. {\bf D23}(1981)558.
\bibitem{dH} E.D'Hoker and D.H.Phong, Rev.Mod.Phys. {\bf 60}(1988)917.
\bibitem{Bir} D.Birmingham, M.Blau, M.Rakowski and G.Thompson, Phys.Rep.
{\bf 209}(1991)129.

\bibitem{det} A.Polyakov and P.B.Wiegmann, Phys.Lett. {\bf 131B}(1983)121,
R.E. Gam\-boa Sarav\'\i , F.A.Schaposnik and J.E.Solom\'\i n,
Nucl.Phys. {\bf
185} (1981)239.
\bibitem{PW} A.Polyakov and P.B.Wiegmann, Phys.Lett. {\bf 141B}(1984)223.

\bibitem{BGRN} L.Brown, G.Goldberg, C.Rim and R.Nepomechie, Phys.Rev.
{\bf D36} (1987)551.
\bibitem{BT2} M.Blau, G.Thompson, Phys.Lett. {\bf 255}(1991)535.

\bibitem{soda} J.Soda, Phys.Let. {\bf 267B}(1991)214.
\bibitem{Ki} P.Killingback, Phys.Lett. {\bf223 B}(1989)357.
\bibitem{BT} M.Blau and G.Thompson, Ann.of Phys (NY) {\bf 205}(1990)130.
\bibitem{Witten2} E.Witten, Commun.Math.Phys. {\bf 141}(1991)153.
\bibitem{CLS} L.Cugliandolo, G.Lozano and F.A.Schaposnik, Phys.Lett.
{\bf 244B}(1990) 249.

\end{thebibliography}
\end{document}